\documentclass[onecolumn,aps,12pt]{revtex4}
\usepackage{amsmath,graphicx}
\usepackage{feynmp}
\usepackage{epsf}
\def\,{\ifmmode\mskip\thinmuskip\else\leavevmode\thinspace\fi}

\newcommand\ba{\begin{eqnarray}}
\newcommand\ea{\end{eqnarray}}
\newcommand{\nn}{\nonumber}
\def\Li#1#2{{\mathrm{Li}}_{#1}\left(#2\right)}

\def\fun#1#2{\lower3.6pt\vbox{\baselineskip0pt\lineskip.9pt
\ialign{$\mathsurround=0pt#1\hfil##\hfil$\crcr#2\crcr\sim\crcr}}}

\begin{document}

\title{The cross sections of the muons and charged pions pairs
production at electron-positron annihilation near the threshold}

\author{Yu.M. Bystritskiy}
\email{bystr@theor.jinr.ru}
\affiliation{\small Joint Institute for Nuclear Research, RU-141980 BLTP JINR,
Dubna, Russia}

\author{G.V. Fedotovich}
\email{fedotovich@inp.nsk.su}
\affiliation{\small Institute of Nuclear Physics, SO RAN, Novosibirsk, Russia}

\author{F.V. Ignatov}
\email{ignatov@inp.nsk.su}
\affiliation{\small Institute of Nuclear Physics, SO RAN, Novosibirsk, Russia}

\author{E.A. Kuraev}
\email{kuraev@theor.jinr.ru}
\affiliation{\small Joint Institute for Nuclear Research, RU-141980 BLTP JINR,
Dubna, Russia}

\date{\today}

\begin{abstract}
The processes of muons (tau) and charged pions pairs production
at electron-positron annihilation with $\mathcal O(\alpha)$ radiative corrections
are considered. The calculation results are presented assuming the
energies of final particles (cms implied) to be
not far significantly from threshold production.
The invariant mass distributions for the muon (tau) and pion pairs are
obtained both for the initial and final state radiation.
Some analytical calculations are illustrated numerically.
The pions were assumed to be point-like objects and scalar QED was
applied for calculation. The QED radiative corrections related
to the final state radiation, additional to the well known Coulomb factor,
are treated near threshold region exactly.
\end{abstract}

\maketitle

\section{Introduction}
The current precision of the evaluation of hadron's contribution to anomalous
magnetic moment of muon is mainly driven by the systematic error of the cross
section of pion pairs production at the region where the total cms
energy of pair does not exceed threshold value significantly. Therefore,
the lowest order radiative corrections (RC) and effects due to the Coulomb
interaction in the final state become essential.

For the purposes of comparisons with experimental data the cases with hard
additional photon must be calculated in frames of PT.
It is a weak point of approaches based on dimensional regularization
methods where the separation of soft and hard photon emission can not be
arranged. One of the motivations of this paper is to calculate these
contributions in frames of traditional QED approach with assigning to
photon small mass and calculate the virtual, soft and hard photon
contributions separately.
In papers \cite{Baier:1965bg} the main characteristics of photon emission
at annihilation $e^+e^-$ to pair of charged particles was investigated.
In papers published in 1983,1985 \cite{Bard,Ber}
the spectra and total cross sections were obtained, but the calculation method
was too complicated.

Below using the invariant integration method we repeat in part those calculations
and obtain the explicit expressions for the spectra distributions on the effective
mass of pair and the corresponding contributions to the total cross
sections due to photon radiation by initial or final particles.
We do not consider the interference of these amplitudes assuming the
experimental set-up to be charge-blind. In this case the interference
contribution to the total cross section is zero.

In section \ref{MuonFSR}, \ref{MuonISR} we consider the final state and initial state
radiation of virtual and real photons in muon pair production process.
In section \ref{PionFSR} similar calculations for the
charged pion pair production (assuming pion to be point-like object) are done.
The results presented in sections \ref{MuonFSR}--\ref{PionFSR}
are in agreement with ones obtained in previous papers
\cite{Drees,Bard,Ber}, but have the form more convenient for
different applications. Some of them concerning initial state radiation
 are new ones.
In section \ref{PionFSR} we also discuss some possibilities of experimental
separation of contribution of initial and final state radiation.
In section \ref{Accuracy} the discussion of accuracy of
results obtained is given.
Whenever possible, the analytical results
are used as a cross-check with ultra relativistic limit.

\section {Final state radiation (FSR) in muon pair production}
\label{MuonFSR}

\begin{figure}[h]
 \vbox to 1.5cm {
 \hspace*{0cm}
 \vspace*{3cm}
 \epsfbox{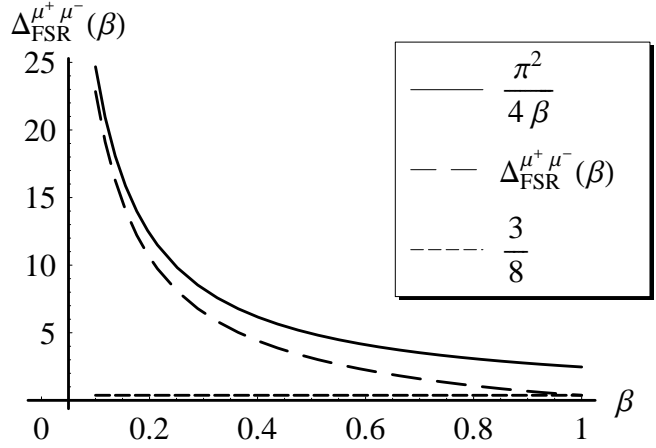}}
 \vbox to 4.cm{}
 \caption{The dependence of quantity $\Delta_{FSR}^{\mu^+\mu^-}(\beta)$ as a
function of $\beta$ for FSR of muons.
See formula (\ref{MuonsFSRTotal1}) and it's asymptotic behavior.}
 \label{MuonsFSRTotalGraph}
 \end{figure}

As well as we are interested in muon effective mass spectrum let
us put the cross section in form:
$$
d\sigma=\frac{1}{8s}\int\sum_{spins}|M|^2d\Gamma.
$$
The summed over spin states matrix element squared can be put in form
(for notations see Fig. \ref{MuonFSRPic}):
\begin{figure}
\begin{fmffile}{MuonFSR}
\begin{fmfgraph*}(150,70)
\fmfleft{Ep,Em}
\fmfright{Mup,Mum}
\fmfright{Gamma}
\fmfv{decor.shape=circle,decor.filled=.5}{MuMuGammaGamma}
\fmf{fermion}{Em,EEGamma,Ep}
\fmf{fermion,width=2}{Mup,MuMuGammaGamma,Mum}
\fmf{photon,label=$q$}{EEGamma,MuMuGammaGamma}
\fmffreeze
\fmf{photon}{MuMuGammaGamma,Gamma}
\fmflabel{$e^-(p_-)$}{Em}
\fmflabel{$e^+(p_+)$}{Ep}
\fmflabel{$\mu^-(q_-)$}{Mum}
\fmflabel{$\mu^+(q_+)$}{Mup}
\fmflabel{$\gamma(k)$}{Gamma}
\end{fmfgraph*}
\end{fmffile}
\caption{Final state radiation corrections to $e^+e^- \to \mu^+\mu^-$ process.}
\label{MuonFSRPic}
\end{figure}
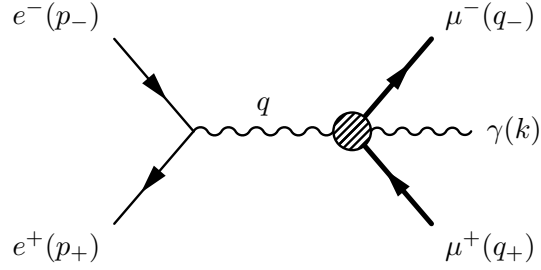
\ba
&&\sum|M|^2=-(4\pi\alpha)^2\frac{1}{s^2}L_{\mu\nu}T^{\mu\nu},
\quad s=(p_++p_-)^2, \nn\\
&&L_{\mu\nu}= Tr\left[\hat{p}_-\gamma_\mu\hat{p}_+\gamma_\nu\right], \quad
T_{\mu\nu}=Tr\left[(\hat{q}_-+M)O_{\mu\eta}(\hat{q}_+-M) \tilde{O}_{\nu\eta}
\right],
\ea
\ba
O_{\mu\nu}=\gamma_\nu\frac{\hat{q}_-+\hat{k}+M}{\chi_-}\gamma_\mu+
\gamma_\mu\frac{-\hat{q}_+-\hat{k}+M}{\chi_+}\gamma_\nu,
\ea
where $\chi_{\pm}=2kq_{\pm}, \quad p_-+p_+=q=q_-+q_++k,\quad q_{\pm}^2=M^2,
\quad p_{\pm}^2=m^2$
and $k^2 = 0.$ $m$, $M$ -- are electron and muon masses correspondingly.
Introducing the energy fractions of final particles we have:
\ba
\nu_{\pm}=\frac{2qq_{\pm}}{s};\quad\nu=\frac{2qk}{s},\quad\nu+\nu_++\nu_-=2,
\nn
\ea
\ba
\int d\Gamma = \int \frac{1}{(2\pi)^5}\frac{d^3q_-}{2E_-}\frac{d^3q_+}{2E_+}
\frac{d^3k}{2\omega}
\delta^4(p_++p_--q_+-q_--k)=\frac{s}{2^7\pi^3}\int_\Delta^{\beta^2}
d\nu\int_{\nu_1}^{\nu_2} d\nu_+, \nn
\ea
\ba
\nu_{1,2}=\frac{1}{2}(2-\nu)\pm\frac{\nu}{2} R(\nu);\qquad
(1-\nu)(1-\nu_-)(1-\nu_+)>\sigma\nu^2, \nn
\ea
\ba
R(\nu)=\sqrt{1-\frac{4\sigma}{1-\nu}}=\sqrt{\frac{\beta^2-\nu}{1-\nu}},
\qquad\beta^2=1-4\sigma,
\qquad \sigma=\frac{M^2}{s}.
\nn
\ea
Due to gauge invariance of tensor $T^{\mu\nu}$ we can write down the following:
$$
\int d\Gamma~T_{\mu\nu}=\frac{1}{3}\left(g_{\mu\nu}-\frac{q_\mu q_\nu}{q^2}
\right)\int d\Gamma~T^\eta_\eta.
$$
Further simplification follows from gauge invariance of initial leptons tensor
$L^{\mu\nu}q_\mu=0$.
Simple calculation gives
\ba
\sum T^\eta_\eta=4\left[\frac{A}{(1-\nu_+)^2}+\frac{B}{1-\nu_+}+C+
(\nu_+\to\nu_-)\right],
\ea
\ba
A&=&-\frac{1}{2}(3-\beta^2)(1-\beta^2),\qquad C=-2; \nn\\
B&=&\frac{1}{\nu}(3-\beta^2)(1+\beta^2) - 2(3-\beta^2)+2\nu.\nn
\ea Integration on the muon energy fraction can be performed using
the expressions: \ba
\int\limits_{\nu_1}^{\nu_2}d\nu_+\left[\frac{1}{(1-\nu_+)^2};
~\frac{1}{1-\nu_+};~1\right]=
\left[\frac{1-\nu}{\nu\sigma}R(\nu);~\ln\frac{1+R(\nu)}{1-R(\nu)};
~\nu R(\nu)\right].
\ea
Distribution on the invariant mass square of muons $m_{\mu\mu}^2=(q_++q_-)^2=s(1-\nu)$
for the case when the energy of hard photon exceeds some value
$\omega>\sqrt{s}\Delta/2, \quad \Delta\ll 1$ has a form
\begin{figure}[h]
 \vbox to 1.5cm {
 \hspace*{0cm}
 \vspace*{3cm}
 \epsfbox{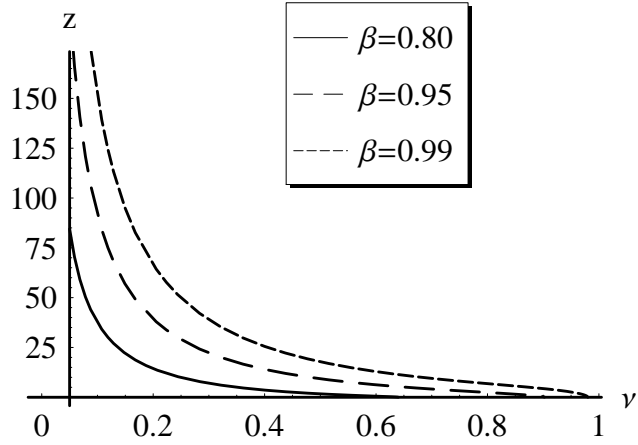}}
 \vbox to 4.cm{}
 \caption{Distribution on $\nu$ for FSR of muons, i.e. the value
$z=\left(2\alpha^3/3s\right)^{-1}(d\sigma^h_{FSR}/d\nu)$
(see (\ref{MuonsFSRHard1})) is shown.}
 \label{MuonsFSRHardGraph}
 \end{figure}
\ba
\frac{d\sigma^h_{FSR}}{d\nu}&=&\frac{2\alpha^3}{3s}
\left[\left[\frac{(1+\beta^2)(3-\beta^2)}{\nu}-2(3-\beta^2)+
2\nu\right]\ln\frac{1+R(\nu)}{1-R(\nu)}-\right.\nn\\
&&\qquad\left.-2\left[\frac{3-\beta^2}{\nu}(1-\nu)+
\nu\right]R(\nu)\right].
\label{MuonsFSRHard1}
\ea
Contribution to the total cross section can be obtained by performing the
integration on invariant muon mass. We use the set of integrals:
\ba
&&\int\limits_\Delta^{\beta^2} R(\nu)\left[\frac{1}{\nu};~1;~\nu\right]d\nu=
\left[-L_\beta+
\beta\ln\frac{4\beta^2}{(1-\beta^2)\Delta};~\beta-\frac{1-\beta^2}
{2}L_\beta;\right.\nn\\
&&\qquad\qquad\qquad\qquad\qquad\left.\beta\frac{3-\beta^2}{4}-
\frac{(3+\beta^2)(1-\beta^2)}{8}L_\beta\right]+O(\Delta),
\ea
\ba
&&\int\limits_\Delta^{\beta^2} \ln\frac{1+R(\nu)}{1-R(\nu)}\left[\frac{1}
{\nu};~1;~\nu\right]d\nu=
\left[L_\beta\ln\frac{1}{\Delta}+ 2\Phi(\beta);~
-\beta+\frac{1}{2}(1+\beta^2)L_\beta;\right.\nn\\
&&\qquad\qquad\qquad\qquad\qquad\left.\frac{1}{16}(3+2\beta^2+3\beta^4)L_\beta-
\frac{3}{8}\beta(1+\beta^2)\right]+O(\Delta),
\ea
with
\ba
L_\beta=\ln\frac{1+\beta}{1-\beta}; \quad
\Phi(\beta)=\Li{2}{1-\beta}-
\Li{2}{1+\beta}-\Li{2}{\frac{1-\beta}{2}}+\Li{2}{\frac{1+\beta}{2}}.
\label{Phi}
\ea
The result is
\ba
\sigma^{e^+e^-\to\mu^+\mu^-\gamma}_{h}&=&\frac{2\alpha}{\pi}\sigma_B(s)
\left[
(\frac{1+\beta^2}{2\beta}L_\beta-1)\ln\frac{1}{\Delta}+\frac{7}{4}-
\ln\frac{4\beta^2}{1-\beta^2}-
\right.\nn\\
&&\qquad\qquad\left.-\frac{3(1+\beta^2)}{8(3-\beta^2)}+
\frac{9-2\beta^2+\beta^4}
{16\beta(3-\beta^2)}L_\beta+
\frac{1+\beta^2}{\beta}\Phi(\beta)\right],
\label{muons_FSR_hard}
\ea
where $\sigma_B(s)=2\pi\alpha^2\beta(3-\beta^2)/(3s)$ is the cross
section of muon pair production in Born approximation.
In the ultra relativistic limit we have
\ba
\left.
\sigma^{e^+e^-\to\mu^+\mu^-\gamma}_{h}\right|_{\beta\to 1}=
\frac{4\pi\alpha^2}{3s}\frac{2\alpha}{\pi}
[(l_\mu-1)\ln\frac{1}{\Delta}-\frac{3}{4}l_\mu+\frac{11}{8}-\xi_2],
\ea
where $l_\mu=\ln(s/M^2)$,~$\xi_2=\pi^2/6$.
This result differs from one given in \cite{kmel}.
The contribution of soft real photons emission with energy
$\omega=\sqrt{k^2+\lambda^2}<\sqrt{s}\Delta/2$,
where $\lambda$ - is "photon mass", is given by:
$$
\sigma_{FSR}^{s}=\sigma_B(s)\left(-\frac{\alpha}{4\pi^2}\right)
\int\frac{d^3k}{\omega}
\left(\frac{q_-}{q_-k}-\frac{q_+}{q_+k}\right)^2,
$$
and performing the standard calculations can be written in form \cite{AkhBer}:
\ba
\sigma_{FSR}^{s}&=&\frac{2\alpha}{\pi}\sigma_B(s)\left[\left(\frac{1+\beta^2}
{2\beta}L_\beta-1\right)\left(\ln\frac{M}{\lambda}+\ln\Delta\right)+
\right.\nn \\
&&\frac{1+\beta^2}{2\beta}\left[\frac{1}{4}L_\beta^2-\Li{2}{\beta}+\Li{2}{-\beta}-
\Li{2}{\frac{1-\beta}{2}}-\right.\nn \\
&&\left.\ln\left(\frac{1+\beta}{2}\right)\ln(1-\beta)+
\frac{1}{2}\ln^2(1+\beta)+
\Li{2}{\frac{1}{2}}+L_\beta\ln\frac{2}{1+\beta}\right]+ \nn \\
&&\left.\ln\left(\frac{1+\beta}{2}\right)+\frac{1-\beta}
{2\beta}L_\beta\right].
\ea
The correction of virtual photon emission include the Dirac and Pauli
formfactors of muon. It has a form  \cite{Schwinger}:
\ba
\sigma^{v}_{FSR}&=&\frac{2\alpha}{\pi}\sigma_B(s)[(1-\frac{1+\beta^2}
{2\beta}L_\beta)\ln\frac{M}{\lambda}- \nn \\
&&1+(\frac{1+\beta^2}{2\beta}-\frac{1}{4\beta})L_\beta+\frac{1+\beta^2}
{2\beta}[2\xi_2-\frac{1}{4}L_\beta^2- \nn \\
&&L_\beta\ln\frac{2\beta}{1+\beta}+\Li{2}{\frac{1-\beta}{1+\beta}}]-
\frac{3(1-\beta^2)}{4\beta(3-\beta^2)}L_\beta].
\ea
The sum of contributions from virtual and soft real photons reads to be:
\ba
\sigma^{v+s}_{FSR}&=&\frac{2\alpha}{\pi}\sigma_B(s)
\left[
\left(\frac{1+\beta^2}
{2\beta}L_\beta-1\right)\ln\Delta - 1 + \ln\frac{1+\beta}{2}+\right.\nn\\
&&
+\left(\frac{3-2\beta+2\beta^2}{4\beta}-\frac{3(1-\beta^2)}
{4\beta(3-\beta^2)}\right)L_\beta+\nn\\
&&\left.
+\frac{1+\beta^2}{2\beta}
\left(-2 \Li{2}{\beta}+2 \Li{2}{-\beta}+\Li{2}{\frac{1+\beta}{2}}-
\Li{2}{\frac{1-\beta}{2}}+3\xi_2\right)
\right].
\ea
In ultra relativistic limit we have:
\ba
\left.\sigma^{v+s}_{FSR}\right|_{\beta\to1}&=&\frac{2\alpha}{\pi}\sigma_B(s)
\left[(l_\mu-1) \ln\Delta - 1 + \frac{3}{4} l_\mu + \xi_2 \right].
\ea
The total sum of contributions from virtual, soft and hard real photons
does not contain photon mass $\lambda$ and
the separation parameter $\Delta$:
\ba
\sigma^{e^+e^-\to\mu^+\mu^-\gamma}_{FSR} = \frac{2\alpha}{\pi} \sigma_B(s)
\Delta_{FSR}^{\mu^+\mu^-}(\beta),
\ea
where
\ba
\Delta_{FSR}^{\mu^+\mu^-}(\beta) &=& \frac{3(5-3\beta^2)}{8(3-\beta^2)}+
\frac{(1-\beta)(33-39\beta-17\beta^2+7\beta^3)}{16\beta(3-\beta^2)}L_\beta+
\nn\\
&+&
3\ln\left(\frac{1+\beta}{2}\right)-2\ln \beta + \frac{1+\beta^2}
{2\beta}F(\beta), \label{MuonsFSRTotal1}\\
F(\beta) &=&
-2 \Li{2}{\beta}+2 \Li{2}{-\beta}
-2 \Li{2}{1+\beta}+2 \Li{2}{1-\beta}\nn\\
&+&3 \Li{2}{\frac{1+\beta}{2}}
-3 \Li{2}{\frac{1-\beta}{2}}+3~\xi_2. \label{FunctionF}
\ea
The quantity $\Delta_{FSR}^{\mu^+\mu^-}(\beta)$ agrees with the result
obtained in \cite{Vol} and presented in Fig. \ref{MuonsFSRTotalGraph}
as a function of $\beta$. This correction in ultra relativistic limit
tends to the value 3/8.
$$
\sigma^{e^+e^-\to\mu^+\mu^-\gamma}_{FSR}|_{\beta\to 1}=
\frac{4\pi\alpha^2}{3s}
\frac{2\alpha}{\pi}\frac{3}{8}=\frac{\alpha^3}{s}.
$$
Cancellation of "large" logarithms $l_\mu=\ln(s/M^2)$ is the consequence
of Kinoshita-Lee-Nauenberg theorem \cite{KLN}.

\section {Initial state radiation (ISR) in muon pair production}
\label{MuonISR}

\begin{figure}
\begin{fmffile}{MuonISR}
\begin{fmfgraph*}(150,70)
\fmfleft{Ep,Em}
\fmftop{Gamma}
\fmfright{Mup,Mum}
\fmfv{decor.shape=circle,decor.filled=.5}{EEGammaGamma}
\fmf{fermion}{Em,EEGammaGamma,Ep}
\fmf{fermion,width=2}{Mup,MuMuGamma,Mum}
\fmf{photon,label=$q$}{EEGammaGamma,MuMuGamma}
\fmffreeze
\fmf{photon}{EEGammaGamma,Gamma}
\fmflabel{$e^-(p_-)$}{Em}
\fmflabel{$e^+(p_+)$}{Ep}
\fmflabel{$\mu^-(q_-)$}{Mum}
\fmflabel{$\mu^+(q_+)$}{Mup}
\fmflabel{$\gamma(k)$}{Gamma}
\end{fmfgraph*}
\end{fmffile}
\caption{Initial state radiation corrections to $e^+e^- \to \mu^+\mu^-$ process.}
\label{MuonISRPic}
\end{figure}
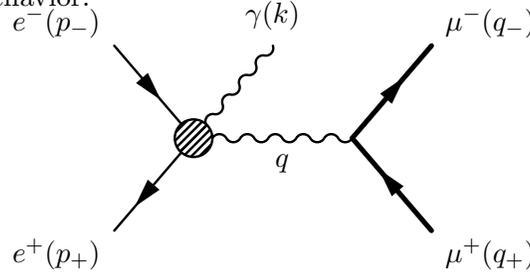

Matrix element of the process of muon pair production with hard photon
radiated from initial state has a form (for notations see Fig. \ref{MuonISRPic}):
\ba
M_{ISR}=\frac{(4\pi\alpha)^{3/2}}{s(1-\nu)}\bar{v}(p_+)
\left[\hat{Q}\frac{\hat{p}_--\hat{k}+m}{-2kp_-}\hat{e}(k)+
\hat{e}(k)\frac{-\hat{p}_++\hat{k}+m}{-2kp_+}\hat{Q}\right]u(p_-),
\ea
with $Q_\eta=\bar{u}(q_-)\gamma_\eta v(q_+)$ is the muon current.
Using the gauge condition for muon current $q^\eta Q_\eta=0$,
~$q=q_++q_-=p_++p_--k$
we have:
\ba
\sum\int Q_\mu(Q_\nu)^* \frac{d^3q_+}{2E_+}\frac{d^3q_-}
{2E_-}\delta^4(q-q_+-q_-)=
D\left(g_{\mu\nu}-\frac{q_\mu q_\nu}{q^2}\right),
\ea
\ba
D=-\frac{2\pi s}{3}\left[\frac{3-\beta^2}{2}-\nu\right]R(\nu),
\qquad q^2=s(1-\nu), \nn
\ea
with notations given above. Using these relations, the calculation of
the summed upon
spin states of matrix element squared is straightforward.
Performing the angular integrations
\ba
\int\limits_{-1}^1dc [\frac{1}{1-\beta_e c};\frac{4m^2}
{s(1-\beta_e c)^2};1]=
[l_e;2;2],\qquad l_e=\ln\frac{s}{m^2},
\qquad \beta_e=\sqrt{1-\frac{4m^2}{s}},
\ea
we obtain the distribution on the muons invariant mass
(see Fig. \ref{MuonsISRHardGraph}):
\begin{figure}[h]
 \vbox to 1.5cm {
 \hspace*{0cm}
 \vspace*{3cm}
 \epsfbox{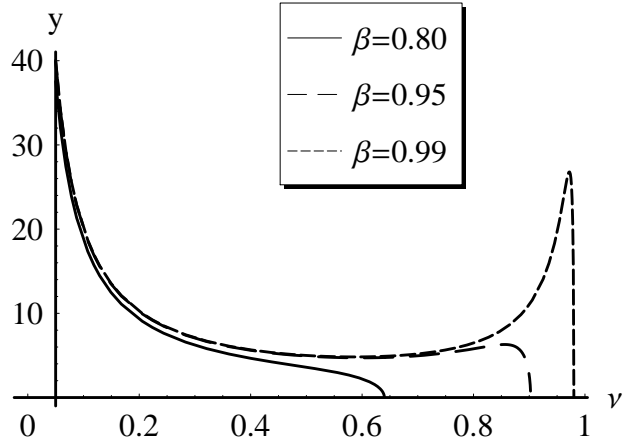}}
 \vbox to 4.cm{}
 \caption{Distribution of muon pairs as a function on $\nu$ for ISR. Vertical axis
represents the quantity
 $y=[(4\alpha^3/3s)(l_e-1)]^{-1}(d\sigma_{ISR}^{h}/d\nu)$ (see (\ref{MuonsISRHard1})), horizontal axis -
the fraction of radiated photon energy $\nu$.}
 \label{MuonsISRHardGraph}
 \end{figure}
\ba
\frac{d\sigma_{ISR}^{h}}{d\nu}=\frac{4\alpha^3}{3s\nu(1-\nu)^2}
[1+(1-\nu)^2](l_e-1)\left(\frac{3-\beta^2}{2}-\nu\right)R(\nu),\quad
\nu > \Delta.
\label{MuonsISRHard1}
\ea
Further integration on the photon energy fraction $\nu$ can be performed using
the set of integrals given above and two additional ones:
$$
\int\limits_0^{\beta^2} R(\nu)[\frac{1}{(1-\nu)^2};\frac{1}{1-\nu}]d\nu
=
[\frac{2\beta^3}{3(1-\beta^2)};-2\beta+L_\beta].
$$
As a result we obtain the cross section due to radiation of hard photon from ISR:
\ba
\sigma_{ISR}^h=\frac{2\alpha}{\pi}\sigma_B(s)(l_e-1)\left[\ln\frac{1}{\Delta}
-\frac{1-3\beta+\beta^3}{\beta(3-\beta^2)}L_\beta - \frac{4}
{3}+2\ln\frac{2\beta}{1+\beta}\right].
\ea
The contribution to the cross section
taking into account the virtual and soft photons to the initial
state is given by:
\ba
\sigma_{ISR}^{s+v}&=&\frac{2\alpha}{\pi}\sigma_B(s)[(l_e-1)\ln\Delta+
\frac{3}{4}l_e-1+\xi_2].
\label{MuonsISRSoftVirt1}
\ea
Let us note that the spectral distribution on invariant mass of
final system have a form consistent with renormalization group
prescriptions, namely one can recognize the kernel of evolution equation
contribution (see (\ref{MuonsISRHard1}), (\ref{MuonsISRSoftVirt1})).

\begin{figure}[h]
 \vbox to 1.5cm {
 \hspace*{0cm}
 \vspace*{3cm}
 \epsfbox{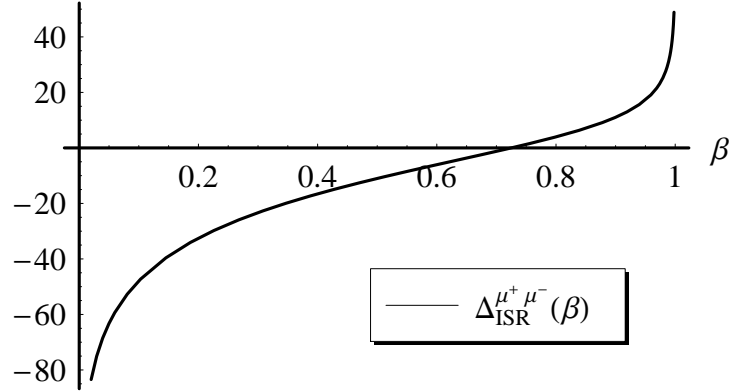}}
 \vbox to 4.cm{}
 \caption{Distribution of muon pairs on $\beta$ for ISR. See formula
(\ref{MuonsISRTotal1}) for the quantity $\Delta_{ISR}^{\mu^+\mu^-}(\beta)$.}
 \label{MuonsISRTotalGraph}
 \end{figure}

Now, we can collect all the terms mentioned above and to write out the
expression for the total cross section due to ISR:
\ba
\sigma_{ISR}^{s+v+h} &=& \frac{2\alpha}{\pi} \sigma_B(s)
\Delta^{\mu^+\mu^-}_{ISR}(\beta), \\
\Delta^{\mu^+\mu^-}_{ISR}(\beta)  &=& (l_e-1)\left[-\frac{1-3\beta+\beta^3}
{\beta(3-\beta^2)}L_\beta
- \frac{4}{3}+2\ln\frac{2\beta}{1+\beta}
\right]+\frac{3}{4}l_e-1+\xi_2. \label{MuonsISRTotal1}
\ea
The dependence of this quantity on muon's velocity $\beta$ is shown in
Fig. \ref{MuonsISRTotalGraph}.
In ultra relativistic limit we have:
\ba
\sigma_{ISR+FSR}^{s+v+h}|_{\beta\to 1}=\frac{8\alpha^3}{3s}
[\frac{1}{2}l_el_\mu-
\frac{1}{2}l_\mu-\frac{7}{12}l_e+\xi_2+\frac{17}{24}],
\ea
which is in agreement with \cite{Bard,Ber}. Leading term
$\sim l_e l_\mu$ is in agreement with the result~\cite{kmel}.

The total cross section contains the so called double-logarithmical
terms ($\sim l_e l_\mu$), which already contradict the renormalization
group predictions (single-logarithmic).

\section {The final state radiation in pion pair production}
\label{PionFSR}
It is worth to remind that the total cross section
$\sigma(e^+e^- \to \pi^+\pi^-)$ with $\mathcal O(\alpha)$ corrections
is required in many subjects of particle physics. Particularly it is required
to determine with a better accuracy the precision of the evaluation of vacuum
polarization effects in photon propagator. The other well known application is
the calculation of the hadronic contribution to the
anomalous magnetic moment of muon $a^{hadr}_{\mu}$~\cite{integr}:
\ba
a^{hadr}_{\mu} = \frac{1}{3}\left(\frac{\alpha}{\pi}\right)^2
\int\limits_{4M^2_\pi}^{\infty}ds~\frac{R(s)K(s)}{s},
\qquad
R(s) = \frac{\sigma^{e^+e^- \to \pi^+\pi^-}(s)}{\sigma^{e^+e^- \to \mu^+\mu^-}(s)}.
\ea
A contribution to this integral coming from high energy region can be
calculated within QCD framework, while for the low energy range the
experimental values R(s) have to be taken as an input~\cite{jeger}.
A numerical evaluation of this integral
in relative unities gives the value $\sim 70$~ppm.

The goal of the new experiment at BNL (E969) is to measure the
anomalous magnetic moment of muon with the relative accuracy of about
$\sim 0.25$~ppm and to improve the previous result~\cite{bnl} by
a factor of two.
It follows that the value $a^{hadr}_{\mu}$ should be calculated
as precisely as possible. In this context the required theoretical
precision of the cross sections with radiative corrections (RC) as well as
the calculation accuracy of the vacuum polarization effects should be
not worse than $\sim 0.2\%$ as it follows from the estimation:
70~ppm$\times 0.2\% \sim $0.14~ppm. This short observation shows
why high precision calculation of the hadronic cross sections are extremely
important.

\subsection{Final state radiation}

\begin{figure}[h]
 \vbox to 1.5cm {
 \hspace*{0cm}
 \vspace*{3cm}
 \epsfbox{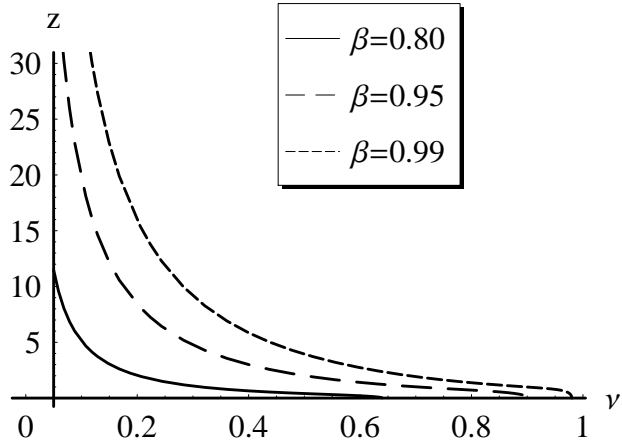}}
 \vbox to 4.cm{}
 \caption{The pion invariant mass distribution on $\nu$ for FSR. The vertical
and horizontal axes represent the value
$z=[(2\alpha^3/3s)]^{-1}(d\sigma^h_{FSR}/d\nu)$ (see (\ref{PionsFSRHard1})) and fraction of photon energy, respectively.}
 \label{PionsFSRHardGraph}
 \end{figure}

As well as it was done for the muons, the contributions with one photon
radiation in the final state can be divided into three separate parts:
virtual, soft and hard. The expression for the virtual photon emission
from final state can be found in~\cite{JHEP} and is given by
\ba
\sigma_v&=&\frac{\alpha}{\pi}\sigma^{\pi^+\pi^-}_B(s)
\biggl[2\ln\frac{M_\pi}{\lambda}
\biggl(1-\frac{1+\beta^2}{2\beta}L_\beta \biggr)-2+
\frac{1+\beta^2}{\beta}L_\beta
\nonumber \\
&+&\frac{1+\beta^2}{\beta}\biggl(-\frac{1}{4}L_\beta^2+
L_\beta\ln\frac{1+\beta}{2\beta}
+2\xi_2+\Li{2}{\frac{1-\beta}{1+\beta}}\biggr)\biggr].
\ea
Here $L_\beta,\lambda,\beta$ were defined above, $\beta$ is a pion velocity
in c.m. frame,
$\sigma^{\pi^+\pi^-}_B(s)=(\pi\alpha^2\beta^3)/(3s)|F_\pi(s)|^2$ is the cross
section production of charged pion pair in the Born approximation,
~$F_\pi(s)$ - pion strong interaction formfactor.
The cross section is due to emission of soft photon when its energy does not
exceed $\Delta\varepsilon$ is given by:
\ba
\label{vs}
\sigma_{FSR}^s&=&\frac{\alpha}{\pi}\sigma^{\pi^+\pi^-}_B(s)
\biggl[2\ln\biggl(\frac{2\Delta\varepsilon}{\lambda}\biggr)
\biggl(\frac{1+\beta^2}{2\beta}L_\beta-1\biggr)+\frac{1}
{\beta}L_\beta \\ \nonumber
&+&\frac{1+\beta^2}{\beta}\biggl(-\frac{1}{4}L_\beta^2
+L_\beta\ln\frac{1+\beta}{2\beta}-\xi_2
+\Li{2}{\frac{1-\beta}{1+\beta}}\biggr)
\biggr], \quad \Delta\varepsilon\ll \varepsilon=\frac{\sqrt{s}}{2}.
\ea
The sum of the contributions from virtual and soft photons
can be presented in convenient way as:
\ba
\sigma_{FSR}^{v+s}=\frac{2\alpha}{\pi}\sigma^{\pi^+\pi^-}_B(s)
\biggl[\biggl(\frac{1+\beta^2}{2\beta}L_\beta-1\biggr)
\ln\Delta+b(s)\biggr],
\ea
where
\begin{gather}
b(s)=-1+\frac{1-\beta}{2\beta}\rho+\frac{2+\beta^2}{\beta}\ln\frac{1+\beta}{2}
+\frac{1+\beta^2}{2\beta}
\biggl[\rho+\xi_2+L_\beta\ln\frac{1+\beta}{2\beta^2}
+2\Li{2}{\frac{1-\beta}{1+\beta}}\biggr],
\nonumber\\
\rho=\ln\frac{4}{1-\beta^2}, \qquad \Delta=\frac{\Delta\varepsilon}
{\varepsilon}.\nonumber
\end{gather}
Calculations similar to ones given above for muons FSR
lead to the pion pair invariant mass distribution ($m_{\pi\pi}^2=s(1-\nu)$,
see Fig.\ref{PionsFSRHardGraph}):
\ba
\frac{\sigma_{FSR}^h}{d\nu}&=&
\frac{2\alpha^3\beta^2}{3s}\left[
\left(\frac{\nu}{\beta^2}-\frac{1-\nu}{\nu}\right)R(\nu)
+\left(\frac{1+\beta^2}{2\nu}-1\right)
\ln\frac{1+R(\nu)}{1-R(\nu)}\right]|F_\pi(s)|^2.
\label{PionsFSRHard1}
\ea
Contribution to the total cross section can be obtained performing the
integration on invariant pion pair mass. In agreement with
\cite{Fpi,Drees} the  relevant contribution has a form:
\ba
\label{hard}
\sigma_{FSR}^h&=&\frac{2\alpha}{\pi}\sigma^{\pi^+\pi^-}_B(s)
\biggl[\ln\frac{1}{\Delta}\biggl(\frac{1+\beta^2}{2\beta}L_\beta-1\biggr)+2
+\frac{3-\beta^2}{4\beta^2}-\frac{(3+\beta^2)(1-\beta^2)}{8\beta^3}L_\beta
\nonumber \\
&-&\ln\frac{4\beta^2}{1-\beta^2}
+\frac{1+\beta^2}{\beta}
\Phi(\beta)\biggr],
\ea
with $\Phi(\beta)$ defined in (\ref{Phi}).
Now we can write down the complete
expression for the total cross section:
\ba
\label{sect}
\sigma^{e^+e^-\to\pi^+\pi^-\gamma}_{FSR} = \frac{2\alpha}{\pi}
\sigma^{\pi^+\pi^-}_B(s) \Delta_{FSR}^{\pi^+\pi^-}(\beta),
\ea
\ba
\Delta_{FSR}^{\pi^+\pi^-}(\beta)&=&\frac{3(1+\beta^2)}{4\beta^2}-2\ln\beta
+3\ln\frac{1+\beta}{2} + \frac{1+\beta^2}{2\beta}F(\beta) \nn\\
&+&\frac{(1-\beta)(-3-3\beta+7\beta^2-5\beta^3)}{8\beta^3}L_\beta,
\label{PionsFSRTotal1}
\ea
\begin{figure}[h]
 \vbox to 1.5cm {
 \hspace*{0cm}
 \vspace*{3cm}
 \epsfbox{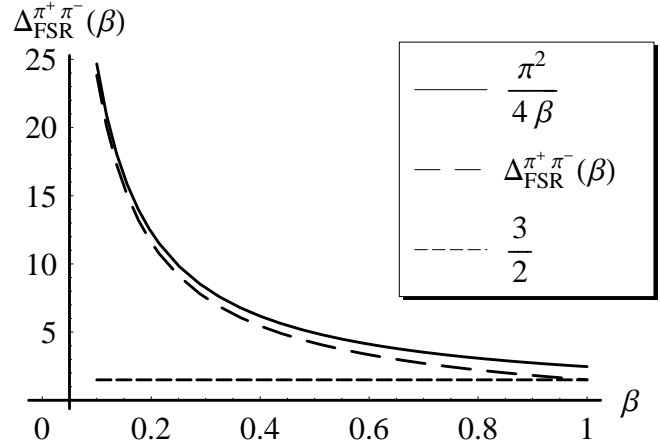}}
 \vbox to 4.cm{}
 \caption{The dependence of quantity $\Delta_{FSR}^{\pi^+\pi^-}(\beta)$
on $\beta$ for FSR of pions. See formula (\ref{PionsFSRTotal1}) and it's
asymptotic behavior.}
 \label{PionsFSRTotalGraph}
 \end{figure}
with the same expression for $F(\beta)$ as in muon case (\ref{FunctionF}).
The factor $\Delta_{FSR}^{\pi^+\pi^-}$ represents
the correction to the Born cross section caused by final state radiation.
In low $\beta$ limit $\Delta_{FSR}^{\pi^+\pi^-}(\beta) \approx \pi^2/4\beta$,
which is the manifestation of Coulomb interaction of pions. There is exactly the same
behavior for $\Delta_{FSR}^{\mu^+\mu^-}(\beta)$ (see (\ref{MuonsFSRTotal1})).
It's well known that in limit $\alpha/\beta \ge 1$ the perturbative
analysis is not valid. The relevant modifications of formulae will be
given in conclusion.

In ultra relativistic limit we have $\Delta_{FSR}^{\pi^+\pi^-}
(\beta \to 1)=3/2$. One can see again, that all
"large" logarithms cancelled out in accordance with
Kinoshita-Lee-Nauenberg theorem. Expression for
$\Delta_{FSR}^{\pi^+\pi^-}(\beta)$ coincide with one obtained
in~\cite{Drees,H}. It is worth noticing that in papers~\cite{Fpi, Drees, H}
the quantity $\Delta_{FSR}^{\pi^+\pi^-}(\beta)$ was presented
without separator $\Delta$ between soft and hard photons.
But for some applications it can be useful to have these two parts
separately.

In order to check experimentally the validity of point-like pions assumption,
which is used in the paper, it's necessary to separate out the
FSR events. Unfortunately we should notice that ISR events 10 times
more probable then the FSR ones. Nevertheless there are at least
two ways to select FSR events and to suppress the ISR background.

Firstly we may consider the region of $\rho$-meson peak left slope,
i.e. $\sqrt{s} < 770$ MeV. In that case the {resonance returning}
mechanism does not take place and the ratio of FSR events increases.

Second way is to throw out the events with pions acollinearity
bigger than some predefined angle, for instance $0.25$ rad
\cite{CMD2}.

\begin{figure}[h]
\begin{minipage}[t]{0.9\textwidth}
 \includegraphics[width=0.9701\linewidth]{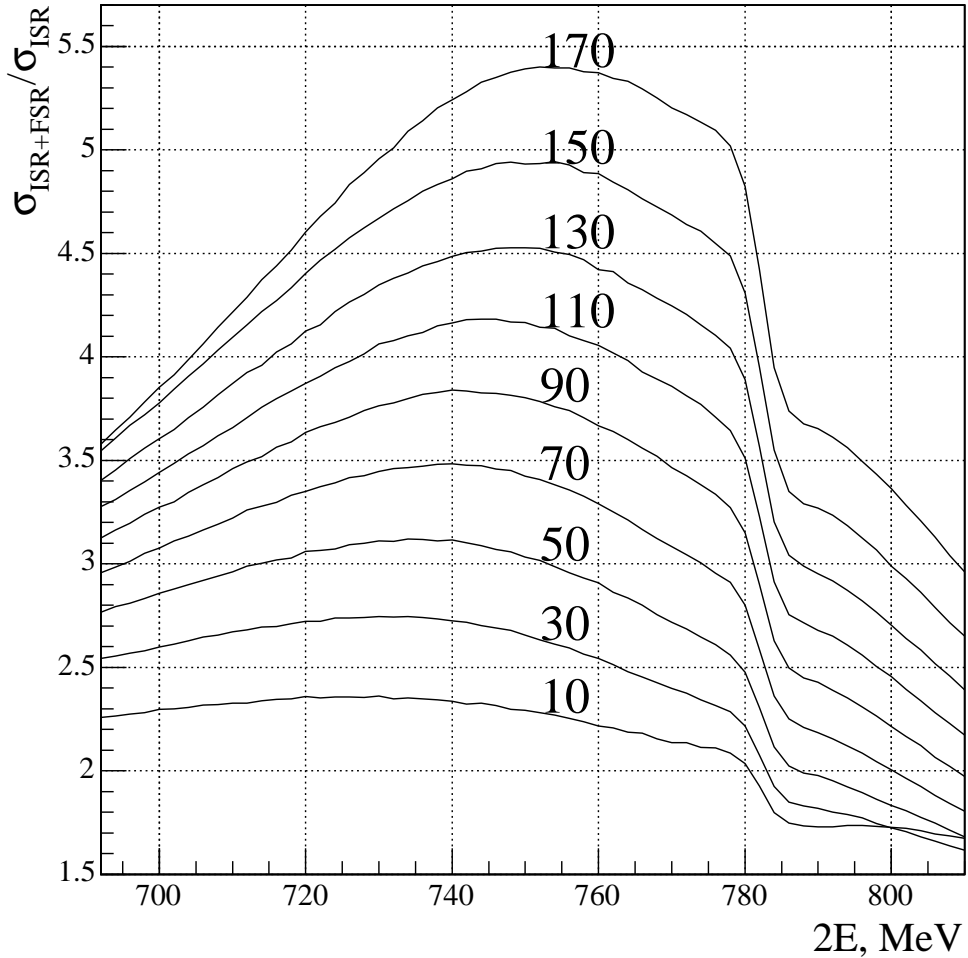}
 \caption{\label{isrfsr_ratio} The ratio of the cross sections with ISR+FSR divided on
the cross section with ISR as a function of energy in c.m. frame.
The different smooth curves represent this ratio {\em vs} threshold photon
energy (in MeV) to be detected.}
 \end{minipage}
 \end{figure}

Figure \ref{isrfsr_ratio} shows the result of modelling of
value $\sigma_{ISR+FSR}/\sigma_{ISR}$ with application of both
FSR separation methods described above. The different curves
correspond to different energy thresholds of emitted photons
($\omega > 10 - 170$ MeV). One can see that the energy range
from 720 to 780 MeV is preferable for our purpose -
if photon energy exceeds $150$ MeV then the ratio
$\sigma_{ISR+FSR}/\sigma_{ISR}$ is about $5$,
this means that the relative admixture of ISR
events is about 20 \% only.
It is worth to notice that the form of spectrum at high photon energies
is just the subject of interest. The comparison of the
simulated spectrum with experimental one can elucidate the discussed problem.

\subsection {Initial state radiation}
\begin{figure}[h]
 \vbox to 1.5cm {
 \hspace*{0cm}
 \vspace*{3cm}
 \epsfbox{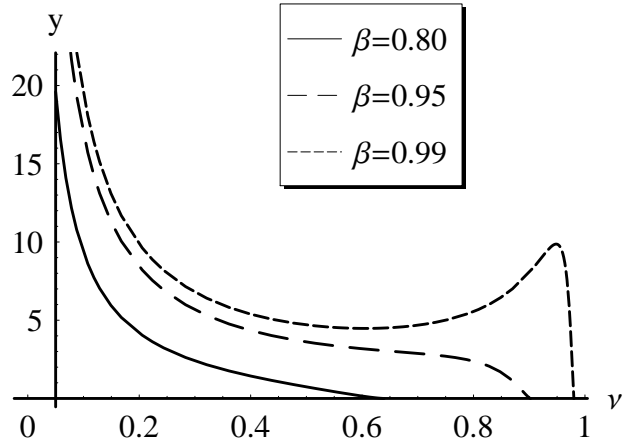}}
 \vbox to 4.cm{}
 \caption{The distribution pion pairs as a function on $\nu$ for ISR. The vertical
axis represents the quantity
 $y=(\alpha^3/3s(l_e-1))^{-1}d\sigma^{e^+e^-\to\pi^+\pi^-\gamma}_{ISR}/d\nu$
(see (\ref{PionsISRHard0})),
horizontal axis - fraction of radiated photon energy.}
 \label{PionsISRHardGraph}
 \end{figure}

Let us consider now the initial state radiation (ISR)
effects in pion pair production. Performing the calculations similar to
the case of muon pair production we have:
\ba
\frac{d\sigma^{e^+e^-\to\pi^+\pi^-\gamma}_{ISR}}{d \nu}
=
\frac{\alpha^3}{3s}
\frac{1+(1-\nu)^2}{(1-\nu)^2\nu} (l_e-1)(\beta^2-\nu)
\sqrt{\frac{\beta^2-\nu}{1-\nu}}|F_\pi(s(1-\nu))|^2,
\label{PionsISRHard0}
\ea
where $q^2=(q_++q_-)^2=s(1-\nu)$. The calculation results are shown in
Fig. \ref{PionsISRHardGraph} with $F_\pi = 1$.
Using integrals presented above we can obtain the following expression
for the cross section with hard photon radiation:
\ba
\sigma^h_{ISR}=
\frac{2\alpha^3\beta^3}{3s}
(l_e-1)
\left\{
    \ln\frac{1}{\Delta}+2\ln\left(\frac{2\beta}{1+\beta}\right)
    -\frac{4}{3}-\frac{1}{2\beta^2}
    +\frac{1-3\beta^2+4\beta^3}{4\beta^3}L_\beta
\right\}, \label{PionsISRHard1}
\ea
where $l_e=\ln(s/m^2)$.
Here we had assumed the pions to be point-like, i.e $F_\pi=1$.
The sum of the contributions of virtual and soft photon emission has a form:
\ba
    &&\sigma_{ISR}^{v+s}=
    \frac{2\alpha}{\pi}\sigma_B^{\pi^+\pi^-}(s)
    \left\{
        (l_e-1)\ln\Delta + \frac{3}{4}l_e - 1+\xi_2
    \right\}.
\ea
The total cross section accounted for initial state radiation can be presented as:
\ba
\label{sect1}
\sigma^{e^+e^- \to \pi^+\pi^-\gamma}_{ISR} =
\frac{2\alpha^3\beta^3}{3s} \Delta_{ISR}^{\pi^+\pi^-}(\beta),
\ea
\ba
\Delta_{ISR}^{\pi^+\pi^-}(\beta)&=&
(l_e-1)\left[2\ln\frac{2\beta}{1+\beta}-\frac{4}{3}-\frac{1}{2\beta^2}+
\frac{1-3\beta^2+4\beta^3}{4\beta^3}L_\beta\right]+\frac{3}{4}l_e-1+\xi_2.
\label{PionsISRTotal1}
\ea
\begin{figure}[h]
 \vbox to 1.5cm {
 \hspace*{0cm}
 \vspace*{3cm}
 \epsfbox{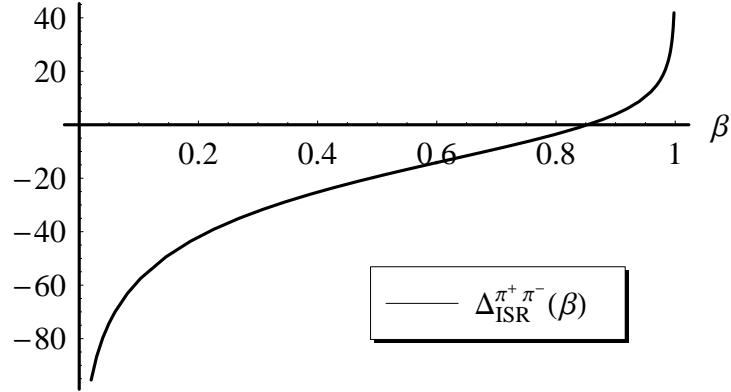}}
 \vbox to 4.cm{}
 \caption{The dependence of the quantity $\Delta_{ISR}^{\pi^+\pi^-}(\beta)$
 (see (\ref{PionsISRTotal1}))
on $\beta$ for ISR.}
 \label{PionsISRTotalGraph}
 \end{figure}
Quantity $\Delta_{ISR}^{\pi^+\pi^-}(\beta)$ as a function of $\beta$ is shown
in Fig. \ref{PionsISRTotalGraph}.
In ultra relativistic limit in point-like approximation for pions we have:
\ba
    \left.
    \sigma^{e^+e^- \to \pi^+\pi^-\gamma}_{ISR+FSR}
    \right|_{\beta \to 1}
    =
    \frac{2\alpha^3}{3s}
    \left\{
        \frac{1}{2}l_e l_\pi - \frac{1}{2}l_\pi+
        \frac{3}{2}l_e + \frac{1}{6}+\xi_2
    \right\},
\ea
where $l_\pi = \ln(s/M_\pi^2)$.
%
%
%
\section{Accuracy estimation}
\label{Accuracy}
The theoretical uncertainties of the cross sections with $\mathcal O (\alpha)$
corrections given above are defined by the unaccounted higher order
corrections and they are estimated to be at $\sim 0.2\%$ level.
Below the main sources of uncertainties which were omitted in the current
formulae are listed:
\begin{itemize}
\item
Weak interactions not considered
here arising from replacement of virtual photon Green function by
$Z$-boson one. It results in
\ba
d\sigma \to d\sigma
\left[ 1 + \mathcal O \left(
\left(\frac{s}{M_Z^2}\right)^2,
\frac{M_\mu^2}{M_Z^2}
\right) \right]
\ea
which for $\sqrt{s} \le 10 GeV$ is of order or smaller
$0.1 \%$ in charge-blind experimental setup, when we can omit the
$\gamma-Z$ interference contribution.
\item
Here we systematically omit the terms of order
$(m/M_\mu)^2$ compared to $1$
\ba
\mathcal O\left( \frac{m^2}{M_\mu^2} \right) \le 0.1 \%.
\ea
\item
The higher orders contributions (not considered here) can be
separated by two classes. First class, leading by large logarithm
$l_e = \ln (s/m^2)$, is connected with ISR:
\ba
&d\sigma \to d\sigma
\left[ 1 + \mathcal O\left(
\left(\alpha/\pi\right)^2 l_e^2,
\left(\alpha/\pi\right)^2 l_e
\right) \right],& \\
&\mathcal O\left( (\alpha/\pi)^2 l_e^2 \right) \sim 0.2 \%,
\qquad
\mathcal O\left(\left({\alpha}/{\pi} \right)^2 l_e \right) \sim 0.01 \%.& \nn
\ea
These kind of contributions can be
taken into account by structure function approach
as it was done in \cite{Kuraev:1985hb}.
\item
Second class is the higher orders contributions connected with FSR which give
\ba
d\sigma
\left[ 1 + \mathcal O\left(\left( \frac{\alpha}{\pi} l_\beta \right)^2\right)
\right],
\qquad
\mathcal O\left(\left( \frac{\alpha}{\pi} l_\beta \right)^2\right) \sim 0.05 \%.
\ea
In ultra relativistic limit $l_\beta \to \ln (s/M_\mu^2)$ they
as well can be taken into account by structure function method.
\end{itemize}
Considering the uncertainty sources mentioned above as independent,
we can conclude that the total systematic error of the cross
sections with $\mathcal O(\alpha)$ RC is less than 0.22~\%.
However we remind that taking into account of higher order contributions
connected with ISR using structure function approach \cite{Kuraev:1985hb}
allows one to decrease the total error down to level 0.05~\%.
$\Delta_{FSR}^{(\mu,\pi)}$ and $\Delta_{ISR}^{(\mu,\pi)}$ are drawn in
figures and one can see that corrections to the Born cross sections
$(2\alpha/\pi)\Delta$ can reach several percents near threshold.

\section{Conclusion}
One of possible applications of formulae given above -- to be used
for normalization purposes at MC simulation.
Our results can be used also for improvement of the calculation accuracy of vacuum
polarization effects
in the virtual photon propagator at low energies not far significantly from
threshold production. This calculation, in one's turn, is required to improve
the precision of the theoretical prediction for anomalous magnetic moment of
muon.

The expressions for the cross sections of $\tau^+\tau^-$ and $K^+K^-$
production are similarly to that for muons and pions. The muon and kaon masses
as well as pion form factor should be replaced in the above expressions by
the tau and kaon ones, respectively.
The cross section being multiplied by the exact Coulomb factor will
interpolate the energy dependence of the cross section from
the threshold production to the relativistic region.

We do not consider C-odd interference in real and virtual photons
emission - it gives zero contribution to the total cross section.
As well we do not consider effects of virtual photon polarization operator
insertion, it can be found in literature \cite{JHEP,Arbuzov:1997je}.

$\Delta_{FSR}^{(\mu,\pi)}$ and $\Delta_{ISR}^{(\mu,\pi)}$ are drawn in
figures and one can see that corrections to Born cross sections
$(2\alpha/\pi)\Delta$ can reach several percents near threshold.

In regions where $\beta \sim \alpha$ formulae must be modified \cite{Vol}.
Taking into account, that
$\Delta^{(i)}(\beta)\sim \pi^2/4\beta$, $\beta\to0$, we must replace

$$
1+\frac{2\alpha}{\pi}\Delta^{(i)}(\beta) \to
\left(1+\frac{2\alpha}{\pi}\left(\Delta^{(i)}(\beta)-
\frac{\pi^2}{4\beta}\right)\right)f(z)
$$
where $ f(z)= z/(1-e^{-z}) $ is the Sommerfeld-Sakharov factor,
$z=(\pi\alpha/\beta)$.
In region there $\beta \ll \alpha$ the formulae must be modified
according to \cite{Vol,Fadin:1991zw}.

\section{Acknowledgments}
We are grateful to A.~Arbuzov, V.~Bytev and V.~Tayursky for valuable discussions.
We also grateful to the RFBR grant 03-02-17077 and INTAS grant 0366 for support.

\end{document}